\documentclass[11pt, onecolumn, draftcls]{IEEEtran}
\usepackage{epsfig, amsmath, amsfonts, amssymb, eucal}%, endfloat}
\newtheorem{thm}{Theorem}%[section]
\newtheorem{cor}{Corollary}
\newtheorem{lem}{Lemma}

\DeclareMathAlphabet{\eurm}{U}{eur}{m}{n}
\DeclareMathAlphabet{\mathbsf}{OT1}{cmss}{bx}{n}% bold sans serif
\DeclareMathAlphabet{\mathssf}{OT1}{cmss}{m}{sl}% slanted sans serif
\DeclareMathAlphabet{\mathcsf}{OT1}{cmss}{sbc}{n}% condensed sans serif

\DeclareSymbolFont{bsfletters}{OT1}{cmss}{bx}{n}  
\DeclareSymbolFont{ssfletters}{OT1}{cmss}{m}{n}
\DeclareMathSymbol{\bsfGamma}{0}{bsfletters}{'000}
\DeclareMathSymbol{\ssfGamma}{0}{ssfletters}{'000}
\DeclareMathSymbol{\bsfDelta}{0}{bsfletters}{'001}
\DeclareMathSymbol{\ssfDelta}{0}{ssfletters}{'001}
\DeclareMathSymbol{\bsfTheta}{0}{bsfletters}{'002}
\DeclareMathSymbol{\ssfTheta}{0}{ssfletters}{'002}
\DeclareMathSymbol{\bsfLambda}{0}{bsfletters}{'003}
\DeclareMathSymbol{\ssfLambda}{0}{ssfletters}{'003}
\DeclareMathSymbol{\bsfXi}{0}{bsfletters}{'004}
\DeclareMathSymbol{\ssfXi}{0}{ssfletters}{'004}
\DeclareMathSymbol{\bsfPi}{0}{bsfletters}{'005}
\DeclareMathSymbol{\ssfPi}{0}{ssfletters}{'005}
\DeclareMathSymbol{\bsfSigma}{0}{bsfletters}{'006}
\DeclareMathSymbol{\ssfSigma}{0}{ssfletters}{'006}
\DeclareMathSymbol{\bsfUpsilon}{0}{bsfletters}{'007}
\DeclareMathSymbol{\ssfUpsilon}{0}{ssfletters}{'007}
\DeclareMathSymbol{\bsfPhi}{0}{bsfletters}{'010}
\DeclareMathSymbol{\ssfPhi}{0}{ssfletters}{'010}
\DeclareMathSymbol{\bsfPsi}{0}{bsfletters}{'011}
\DeclareMathSymbol{\ssfPsi}{0}{ssfletters}{'011}
\DeclareMathSymbol{\bsfOmega}{0}{bsfletters}{'012}
\DeclareMathSymbol{\ssfOmega}{0}{ssfletters}{'012}

\begin{document}

\title{Joint Transmission and State Estimation: A Constrained Channel Coding Approach}
\author{Wenyi Zhang, {\it Senior Member IEEE}, Satish Vedantam, and Urbashi Mitra, {\it Fellow IEEE}
\thanks{W. Zhang is with Department of Electronic Engineering and Information Science, University of Science and Technology of China, Hefei 230027, China (E-mail: {\tt wenyizha@ustc.edu.cn}). S. Vedantam was with the Ming Hsieh Department of Electrical Engineering, University of Southern California, Los Angeles 90089, USA, and is now with R\&D Department, Bloomberg LP, 731 Lexington Avenue, New York, NY 10017, USA (E-mail: {\tt vedantam@gmail.com}). U. Mitra is with the Ming Hsieh Department of Electrical Engineering, University of Southern California, Los Angeles 90089, USA (E-mail: {\tt ubli@usc.edu}). This work has been supported in part by ONR through contract N-000140410273, NSF through contracts ITR CCF-0313392 and OCE 0520324, an Annenberg Fellowship and by the University of Southern California. Preliminary results of this work have been presented in part in \cite{zhang08:isit}.}}

\maketitle

%%%%%%%%%%%%%%%%%%%%%%%%%%%%%%%%%%%%%%%%%%%%%%%%%%%%%%%%%%%%%%%%%%%%%%
\begin{abstract}
A scenario involving a source, a channel, and a destination, where the destination is interested in {\em both} reliably reconstructing the message transmitted by the source and estimating with a fidelity criterion the state of the channel, is considered. The source knows the channel statistics, but is oblivious to the actual channel state realization. Herein it is established that a distortion constraint for channel state estimation can be reduced to an additional cost constraint on the source input distribution, in the limit of large coding block length. A newly defined capacity-distortion function thus characterizes the fundamental tradeoff between transmission rate and state estimation distortion. It is also shown that non-coherent communication coupled with channel state estimation conditioned on treating the decoded message as training symbols achieves the capacity-distortion function. Among the various examples considered, the capacity-distortion function for a memoryless Rayleigh fading channel is characterized to within $1.443$ bits at high signal-to-noise ratio. The constrained channel coding approach is also extended to multiple access channels, leading to a coupled cost constraint on the input distributions for the transmitting sources.
\end{abstract}
\begin{keywords}
capacity distortion function, channel state estimation, estimation cost, input constraint, multiple access, Rayleigh fading
\end{keywords}

%%%%%%%%%%%%%%%%%%%%%%%%%%%%%%%%%%%%%%%%%%%%%%%%%%%%%%%%%%%%%%%%%%%%%%
\section{Introduction}
\label{sec:intro}

In this paper, we consider the problem of joint information transmission and channel state estimation over a channel with a random time-varying channel state. The objective is to have the destination recover both the message transmitted from the source and the states of the channel over which the message is transmitted, under the presumption that the random channel state is available to {\em neither} the source nor the destination, save its statistics. The problem setting is relevant for general situations where, besides communication, the destination needs to identify intrusions (as in secret communication systems \cite{lee01:ssp}) or interference (as in dynamic spectrum access systems \cite{haykin05:jsac}), to monitor the environment (as in underwater acoustic/sonar applications \cite{stojanovic96:joe}), and others. For example, in active sonar systems, a source can transmit a signal, which experiences Rayleigh fading that is a function of the reflected target \cite{urick83:book}. Thus, the received signal at the receiver is of the form:
\begin{eqnarray*}
Y &=& SX + Z
\end{eqnarray*}
where $X$ is the transmitted signal, $S$ is the information about the target as revealed in a Rayleigh fading channel and $Z$ is channel noise. Extension of our results to the parallel channel enables the consideration of multistatic sonar \cite{coraluppi05:aes}.

%For future cognitive radio applications, one can envision designing secondary user signals that cause limited distortion to primary user signals \cite{your paper}:
%\begin{eqnarray*}
%Y & = & X + S + Z
%\end{eqnarray*}
%where $X$ is the secondary user signal, $S$ is the primary user signal and $Z$ is channel noise.  The secondary user estimates the primary user's signal in order to ensure limited interference to the primary user.  An alternative view is that a primary receiver, receiving the primary user signal $X$, wishes to employ a particular transmission band assuming the interference $S$ is below a certain level; however, it must sense the channel accurately.  Thus the receiver communicates and senses simultaneously, rather than disjointly \cite{favorite specturm sensing paper} ensuring efficient use of resources.
In contrast to much prior work, we consider both information transmission and channel state estimation. In the literature, channel state estimation has long been studied with the goal of facilitating information transmission, versus as a separate goal in itself; see, {\it e.g.}, \cite{hassibi03:it}. The channel state estimation therein is for information transmission only, and does not compete for resources with the data transmission as we consider in this work.

The problem formulation in \cite{sutivong02:isit, sutivong05:it, kim08:it} bears similarity to that we consider: the destination is interested in both information transmission and channel state estimation. However, a critical distinction that differentiates our work from this line of prior work, is the fact that in those works the channel state is assumed to be non-causally known at the source \cite{sutivong02:isit, sutivong05:it, kim08:it} and thus can be exploited for encoding the message. In our formulation, neither the source nor the destination has {\it a priori} knowledge of the channel state, except its statistics. Consequently, the solution for our problem and those of \cite{sutivong02:isit, sutivong05:it, kim08:it} are fundamentally different, as will be elaborated upon in the paper.  However, since submission of our paper, a work \cite{chiru10} that unifies our scenario with that of \cite{sutivong02:isit, sutivong05:it} has been presented; thus connecting the case of non-causal knowledge of state at the transmitter with the case of both the transmitter and the receiver being completely oblivious of channel state as we examine herein.

Intuitively, an inherent tradeoff exists between a channel's capability to transfer information and its capability to reveal state. Information transmission is accomplished by exercising random channel inputs, thereby increasing the randomness of the channel outputs and thus reducing the destination's ability to estimate channel states. Channel state estimation, in contrast, suggests that the source transmit deterministic channel inputs, limiting any information transmission through the channel. We quantitatively characterize such a fundamental tension in this paper.

We show that the optimal transmission rate versus state estimation distortion can be formulated as a constrained channel coding problem, with
the channel input distribution constrained by an average cost constraint wherein which we associate with each input symbol an ``estimation cost''. The problem of designing the optimal source then reduces to selecting codebooks which meet the estimation cost constraint, and the optimal tradeoff between transmission rate and state estimation distortion is characterized by a function termed the {\it capacity-distortion function}. Furthermore, we show that non-coherent communication coupled with channel state estimation conditioned on treating the decoded message as training symbols achieves the capacity-distortion function. We later extend the basic idea to two-user multiple access channels (MAC) with channel state estimation at the destination, and characterize the capacity region-distortion function for that scenario. The channel state estimation constraint again leads to an additional estimation cost constraint on the source distribution; however, this cost constraint is in contrast to conventional MAC in that, here, the estimation cost constraint is a coupled constraint for the two sources, as opposed to the separate input cost constraints such as an average power constraint at each of the sources. Thus, having specified the estimation cost constraint, the sources collaboratively optimize their input distributions, even when there are separate additional cost constraints.

The rest of this paper is organized as follows. Section \ref{sec:problem-formulation} describes the basic channel model with discrete alphabets, and formulates the problem of characterizing the capacity-distortion function. Section \ref{sec:achievability} gives the capacity-distortion function, and establishes its achievability, through formulating the constrained channel coding problem. Section \ref{sec:converse} proves the converse part of the capacity-distortion function.
Section \ref{sec:continuous} extends the results of the previous two, by considering an average cost constraint to the channel inputs in addition to the state estimation constraint.
Section \ref{sec:example} illustrates the application of the capacity-distortion function through several examples, including characterizing the capacity-distortion function for a memoryless Rayleigh fading channel within $1.443$ bits at high signal-to-noise ratio (SNR). Section \ref{sec:mac} establishes the capacity-distortion region for two-user MAC with channel state estimation. Finally, Section \ref{sec:conclusion} concludes the paper.

%%%%%%%%%%%%%%%%%%%%%%%%%%%%%%%%%%%%%%%%%%%%%%%%%%%%%%%%%%%%%%%%%%%%%%
\section{Basic Problem Formulation}
\label{sec:problem-formulation}

In this section, we formulate the joint information transmission and channel state estimation problem. For simplicity, we focus on channels with discrete alphabets.

{\it Message:} An index $m$ uniformly selected among $\mathcal{M} = \{1, 2, ..., |\mathcal{M}|\}$.

{\it Channel input:} A symbol $x$ taken from a finite input alphabet $\mathcal{X} = \{a^{(1)}, a^{(2)}, \ldots, a^{(|\mathcal{X}|)}\}$.

{\it Channel output:} A symbol $y$ taken from a finite output alphabet $\mathcal{Y} = \{b^{(1)}, b^{(2)}, \ldots, b^{(|\mathcal{Y}|)}\}$.

{\it Channel state:} A symbol $s$ taken from a finite state alphabet $\mathcal{S} = \{c^{(1)}, c^{(2)}, \ldots, c^{(|\mathcal{S}|)}\}$. For each channel use, the state is a random variable $S$ which has a probability mass function (PMF) $P_S(s)$. Over any $n$ consecutive channel uses, the channel state sequence $S^n$ is memoryless, $P(s^n) = \prod_{i = 1}^n P_S(s_i)$.

{\it Channel:} A collection of probability transition matrices each of which specifies the conditional probability distribution under a fixed channel state; that is, $P(b^{(j)}|a^{(i)}, c^{(k)})$ represents the probability of output $y = b^{(j)} \in \mathcal{Y}$ occurring given input $x = a^{(i)} \in \mathcal{X}$ and state $s = c^{(k)} \in \mathcal{S}$, for any $1 \leq i \leq |\mathcal{X}|$, $1 \leq j \leq |\mathcal{Y}|$ and $1 \leq k \leq |\mathcal{S}|$. With $n$ consecutive channel uses, the channel transitions are mutually independent, characterized by $\prod_{i = 1}^n P(y_i|x_i, s_i)$ for output $(y_1, \ldots, y_n) \in \mathcal{Y}^n$ occurring given input $(x_1, \ldots, x_n) \in \mathcal{X}^n$ and state $(s_1, \ldots, s_n) \in \mathcal{S}^n$.

{\it Distortion:} For any two channel states, the distortion is a deterministic function, $d: \mathcal{S} \times \mathcal{S} \mapsto \mathbb{R}^+\cup \{0\}$. It is further assumed that $d(\cdot, \cdot)$ is bounded, {\it i.e.}, $d(c^{(i)}, c^{(j)}) \leq \bar{D} < \infty$ for any $1\leq i, j \leq |\mathcal{S}|$. For any two length-$n$ state sequences $(s_1, \ldots, s_n), (s'_1, \ldots, s'_n) \in \mathcal{S}^n$, the average distortion is the average of the pairwise distortions, $(1/n) \sum_{i = 1}^n d(s_i, s'_i)$.

{\it Coding framework:} For each coding block length $n$, an $(|\mathcal{M}|, n)$-code is described by the following components.
\begin{itemize}
\item {\it Encoder:} A deterministic function, $f_n: \mathcal{M} \mapsto \mathcal{X}^n$. Denote the codewords by $x^n(1), \ldots, x^n(|\mathcal{M}|)$, where $x^n(m) = f_n(m)$ for each $m$.
\item {\it Decoder:} A deterministic function, $g_n: \mathcal{Y}^n \mapsto \mathcal{M}$.
\item {\it State estimator:} A deterministic function, $h_n: \mathcal{Y}^n \mapsto \mathcal{S}^n$. We denote $\hat{S}^n = h_n(Y^n)$ as the estimated channel states.
\end{itemize}

{\it Probability of error for information transmission:} We consider the average probability of error, which is defined as
\begin{eqnarray}
P_e^{(n)} = \frac{1}{|\mathcal{M}|} \sum_{m \in \mathcal{M}}\mathbf{Pr}\left[g_n(Y^n) \neq m| X^n = f_n(m) \right],
\end{eqnarray}
where $Y^n$ is induced by the channel input vector $x^n = f_n(m)$ and the channel state vector $S^n$ according to the channel transition probability distributions.

{\it Distortion for channel state estimation:} We consider the average distortion, which is defined as
\begin{eqnarray}
\label{eqn:distortion-def}
\bar{d}^{(n)} = \frac{1}{|\mathcal{M}|} \sum_{m \in \mathcal{M}} \mathbf{E}\left[\left.\frac{1}{n}\sum_{i = 1}^n d(S_i, \hat{S}_i) \right| X^n = f_n(m)\right],
\end{eqnarray}
where the expectation is over the conditional joint distribution of $(S^n, Y^n)$ conditioned by the message $m \in \mathcal{M}$, noting that $\hat{S}^n$ is determined by $Y^n$.

{\it Achievable transmission-state estimation tradeoff:} A pair $(R, D)$, denoting a transmission rate and a state estimation distortion is said to be achievable if there exists a sequence of $(\left\lceil e^{nR} \right\rceil, n)$-codes, indexed by $n = 1, 2, \ldots$, such that $\lim_{n \rightarrow \infty} P_e^{(n)} = 0$, and $\limsup_{n \rightarrow \infty} \bar{d}^{(n)} \leq D$.

{\it Capacity-distortion function:} For every $D \geq 0$, the capacity-distortion function $C(D)$ is the supremum of rates $R$ such that $(R, D)$ is an achievable transmission-state estimation tradeoff.

The central problem in this paper is to characterize $C(D)$.

%%%%%%%%%%%%%%%%%%%%%%%%%%%%%%%%%%%%%%%%%%%%%%%%%%%%%%%%%%%%%%%%%%%%%%
\section{The Capacity-Distortion Function and Proof of Achievability}
\label{sec:achievability}

In this section, we present the capacity-distortion function and establish its achievability.

To characterize $C(D)$, we define the following minimal conditional distortion (or estimation cost) function for each channel input symbol $x \in \mathcal{X}$:
\begin{eqnarray}
\label{eqn:d-ast}
d^\ast(x) = \min_{h: \mathcal{X} \times \mathcal{Y} \mapsto \mathcal{S}} \mathbf{E}\left[d(S, h(X, Y)) | X = x \right],
\end{eqnarray}
where the function $h: \mathcal{X} \times \mathcal{Y} \mapsto \mathcal{S}$ is an one-shot estimator, and the expectation is over the conditional joint distribution of $(S, Y)$ conditioned upon $X = x$, namely,
\begin{eqnarray*}
&&\mathbf{Pr}[S = s, Y = y|X = x]\\
 &=& \mathbf{Pr}[Y = y|X = x, S = s] \mathbf{Pr}[S = s|X = x]\\
 &=& P(y|x, s) P_S(s).
\end{eqnarray*}
Note that $h$ maps a pair of channel input and channel output to a channel state. We denote the function $h$ that attains $d^\ast(\cdot)$ by $h^\ast(\cdot, \cdot)$. When there are more than one one-shot estimators that attain $d^\ast(x)$, an arbitrary one is selected.

The capacity-distortion function is given by the following theorem.
\begin{thm}
\label{thm:c-d}
The capacity-distortion function for the problem considered in Section \ref{sec:problem-formulation} is
\begin{eqnarray}
\label{eqn:c-d}
C(D) = \max_{P_X \in \mathcal{P}_D} I(X; Y),
\end{eqnarray}
where
\begin{eqnarray}
\label{eqn:p-d-set}
\mathcal{P}_D = \left\{P_X: \sum_{x \in \mathcal{X}} P_X(x) d^\ast(x) \leq D \right\}.
\end{eqnarray}
\end{thm}

Inspecting Theorem \ref{thm:c-d}, we see that $d^\ast(x)$ serves as an ``estimation cost'' due to signaling with $x \in \mathcal{X}$. Hence $\mathcal{P}_D$, as defined by (\ref{eqn:p-d-set}), specifies an average cost constraint which regulates the input distribution so that the signaling is ``estimation-efficient''. Note that here the channel transition probability is marginalized over the channel state $S$, {\it i.e.}, $\mathbf{Pr}[Y  = y|X = x] = \sum_{s \in \mathcal{S}} P_S(s) P(y|x, s)$.

Before providing the proof of Theorem \ref{thm:c-d}, we first summarize a few useful properties of $C(D)$ in Corollary \ref{cor:cd-property}.
\begin{cor}
\label{cor:cd-property}
The capacity-distortion function $C(D)$ in Theorem \ref{thm:c-d} has the following properties:
\begin{enumerate}
\item $C(D)$ is defined for all $D \geq d_{\min} = \min_{x\in \mathcal{X}} d^\ast(x)$.
\item $C(D)$ is a non-decreasing concave function of $D$ for all $D \geq d_{\min}$.
\item $C(D)$ is a continuous function of $D$ for all $D > d_{\min}$.
\item If $d_{\min}$ is achieved by a unique $x \in \mathcal{X}$, $C(d_{\min}) = 0$.
\item $C(D) = C(\infty)$, for all $D \geq d_{\max} = \max_{x\in \mathcal{X}} d^\ast(x)$, where $C(\infty)$ is the unconstrained channel capacity.
\end{enumerate}
\end{cor}
Property 2) is established in the converse proof in Section \ref{sec:converse}, and 3) is a direct consequence of Property 2). Properties 1), 4), and 5) are straightforward and thus provided without proof.

In the remainder of this section, we prove the achievability part of Theorem \ref{thm:c-d}.

{\it Proof of achievability:} The transmission part of the achievability proof closely follows the standard channel coding theorem. We fix a distribution $P_X \in \mathcal{P}_D$, and generate a $(\left\lceil e^{nR} \right\rceil, n)$-code at random according to the constant composition of $P_X$ (see, {\it e.g.}, \cite{csiszar98:it}), for $R < I(X; Y)$. The channel coding theorem for constant composition codes ensures that there exists a sequence of $(\left\lceil e^{nR} \right\rceil, n)$-codes which achieves $\lim_{n \rightarrow \infty} P_e^{(n)} = 0$. For each coding block length $n$, we can partition the output space $\mathcal{Y}^n$ into $|\mathcal{M}| = \left\lceil e^{nR} \right\rceil$ disjoint subsets $\mathcal{D}^{(n)}_1, \mathcal{D}^{(n)}_2, \ldots, \mathcal{D}^{(n)}_{|\mathcal{M}|}$, and decode the message index as $g_n(y^n) = \hat{m}$ if the channel output $y^n$ belongs to $\mathcal{D}^{(n)}_{\hat{m}}$.

The next step in the achievability proof concerns the state estimation. %We denote for any sufficiently large coding block length $n$ the decoder achieves
%\begin{eqnarray}
%\label{eqn:decode-highly-correct}
%\epsilon^{(n)}_m = \mathbf{Pr}\left[g_n(Y^n) \neq m| X^n = f_n(m) \right] < \epsilon,
%\end{eqnarray}
%for every message $m \in \mathcal{M}$.
After decoding, the destination then re-encodes the decoded message, to form
\begin{eqnarray}
\hat{X}^n = f_n(g_n(Y^n)).
\end{eqnarray}
The state estimator chooses the state estimator $h_n$ to compute the channel state estimates according to the following:
\begin{eqnarray}
\hat{S}^n = h_n^\ast(Y^n) = \left[h^\ast(\hat{X}_1, Y_1), \ldots, h^\ast(\hat{X}_n, Y_n)\right]^T.
\end{eqnarray}

Now consider the average distortion incurred by the state estimation procedure described above. For every coding block length $n$, for every $m \in \mathcal{M}$, the average distortion is given by
\begin{eqnarray}
\label{eqn:distortion-bound}
\bar{d}^{(n)}_m &=& \mathbf{E}\left[\left.\frac{1}{n}\sum_{i = 1}^n d(S_i, \hat{S}_i) \right| X^n = f_n(m)\right]\nonumber\\
&=& \sum_{(s^n, y^n) \in \mathcal{S}^n \times \mathcal{Y}^n} \mathbf{Pr}[S^n = s^n, Y^n = y^n|X^n = f_n(m)] \frac{1}{n}\sum_{i = 1}^n d(S_i, \hat{S}_i)\nonumber\\
&=& \sum_{(s^n, y^n) \in \mathcal{S}^n \times \mathcal{D}_m^{(n)}} \mathbf{Pr}[S^n = s^n, Y^n = y^n|X^n = f_n(m)] \frac{1}{n}\sum_{i = 1}^n d(S_i, \hat{S}_i) + \nonumber\\
&&\quad \sum_{(s^n, y^n) \in \mathcal{S}^n \times \left(\mathcal{Y}^n \backslash \mathcal{D}_m^{(n)}\right)} \mathbf{Pr}[S^n = s^n, Y^n = y^n|X^n = f_n(m)] \frac{1}{n}\sum_{i = 1}^n d(S_i, \hat{S}_i)\nonumber\\
&\leq& \sum_{(s^n, y^n) \in \mathcal{S}^n \times \mathcal{D}_m^{(n)}} \mathbf{Pr}[S^n = s^n, Y^n = y^n|X^n = f_n(m)] \frac{1}{n}\sum_{i = 1}^n d(S_i, \hat{S}_i) + \epsilon^{(n)}_m \bar{D}\nonumber\\
&=& \sum_{(s^n, y^n) \in \mathcal{S}^n \times \mathcal{D}_m^{(n)}} \mathbf{Pr}[S^n = s^n, Y^n = y^n|X^n = f_n(m)] \frac{1}{n}\sum_{i = 1}^n d(S_i, h^\ast(x_i(m), Y_i)) + \epsilon^{(n)}_m \bar{D}\nonumber\\
&\leq& \sum_{(s^n, y^n) \in \mathcal{S}^n \times \mathcal{Y}^n} \mathbf{Pr}[S^n = s^n, Y^n = y^n|X^n = f_n(m)] \frac{1}{n}\sum_{i = 1}^n d(S_i, h^\ast(x_i(m), Y_i)) + \epsilon^{(n)}_m \bar{D}\nonumber\\
&=& \mathbf{E}\left[\left. \frac{1}{n}\sum_{i = 1}^n d(S_i, h^\ast(X_i, Y_i))\right | X_i = x_i(m)\right] + \epsilon^{(n)}_m \bar{D},
\end{eqnarray}
where the expectation is over the distribution of $(S^n, Y^n)$, and we define
\begin{eqnarray}
\epsilon^{(n)}_m = \mathbf{Pr}\left[g_n(Y^n) \neq m| X^n = f_n(m) \right].
\end{eqnarray}
At this point, it directly follows, from the linearity of expectation and the definition of $d^\ast(\cdot)$ in (\ref{eqn:d-ast}), that
\begin{eqnarray}
\label{eqn:distortion-bounding-3}
\bar{d}^{(n)}_m \leq \frac{1}{n} \sum_{i = 1}^n d^\ast(x_i(m)) + \epsilon^{(n)}_m \bar{D}.
\end{eqnarray}
%On the other hand, it is also evident that $\bar{d}^{(n)}_m$ can be lower bounded by
%\begin{eqnarray}
%\label{eqn:d-bar-lb}
%\bar{d}^{(n)}_m \geq \frac{1}{n} \sum_{i = 1}^n d^\ast(x_i).
%\end{eqnarray}

As we average the per-message average distortions $\bar{d}^{(n)}_m$ over $\mathcal{M}$ following (\ref{eqn:distortion-def}), the average distortion $\bar{d}^{(n)}$ is
\begin{eqnarray}
\label{eqn:distortion-bounding-4}
\bar{d}^{(n)} &=& \frac{1}{|\mathcal{M}|} \sum_{m \in \mathcal{M}} \bar{d}^{(n)}_m\nonumber\\
&\leq& \frac{1}{n|\mathcal{M}|} \sum_{m \in \mathcal{M}} \sum_{i = 1}^n d^\ast(x_i(m)) + \bar{D} \frac{1}{|\mathcal{M}|}\sum_{m \in \mathcal{M}} \epsilon^{(n)}_m\nonumber\\
&=& \frac{1}{n|\mathcal{M}|} \sum_{m \in \mathcal{M}} \sum_{i = 1}^n d^\ast(x_i(m)) + \bar{D} P_e^{(n)}.
\end{eqnarray}

Recall that the codebook is generated according to a constant composition of input distribution $P_X$. Therefore, by letting $n \rightarrow \infty$, we have from (\ref{eqn:distortion-bounding-4})
\begin{eqnarray}
\label{eqn:distortion-asymptote}
\limsup_{n \rightarrow \infty} \bar{d}^{(n)} \leq \sum_{x\in \mathcal{X}} P_X(x) d^\ast(x) \leq D,
\end{eqnarray}
where the last inequality is from the fact that $P_X$ belongs to $\mathcal{P}_D$. Finally, by optimizing the possible input distribution $P_X$ over $\mathcal{P}_D$, we establish the achievability of $C(D)$.

%%%%%%%%%%%%%%%%%%%%%%%%%%%%%%%%%%%%%%%%%%%%%%%%%%%%%%%%%%%%%%%%%%%%%%
\section{Proof of Converse}
\label{sec:converse}

In this section, we prove that for every achievable rate-distortion pair $(R, D)$, $R < C(D)$ holds.

{\it Proof:} For an arbitrarily chosen achievable rate-distortion pair $(R, D)$, consider a $(\left\lceil e^{nR}\right\rceil, n)$-code that achieves it. Applying Fano's inequality as in the standard channel coding theorem \cite[Ch. 7, Sec. 9]{cover06:book}, we have
\begin{eqnarray}
R \leq \frac{1}{n} I(X^n; Y^n) + P_e^{(n)} R + \frac{1}{n}.
\end{eqnarray}
Here the distribution of $X^n$ is induced by the uniformly selected message, and the distribution of $Y^n$ is correspondingly induced by $X^n$ and $S^n$. Since the channel is memoryless, through standard bounding steps, $R$ is upper bounded by
\begin{eqnarray}
\label{eqn:R-bound-standard}
R \leq \frac{1}{n}\sum_{i = 1}^n I(X_i; Y_i) + P_e^{(n)} R + \frac{1}{n}.
\end{eqnarray}
From the definition of $C(D)$ in Theorem \ref{thm:c-d}, (\ref{eqn:R-bound-standard}) further leads to
\begin{eqnarray}
\label{eqn:R-bound-c-d}
R \leq \frac{1}{n} \sum_{i = 1}^n C\left(\sum_{x \in \mathcal{X}}P_{X_i}(x) d^\ast(x)\right) + P_e^{(n)} R + \frac{1}{n}.
\end{eqnarray}

At this point, we note that $C(D)$ is a non-decreasing and concave function of $D$. The non-decreasing property is clear because $\mathcal{P}_{D_1} \subseteq \mathcal{P}_{D_2}$ for arbitrary $D_1 \leq D_2$. To see the concavity property, denote the input distribution that achieves $C(D_j)$ ($j = 1, 2$) by $P_X^{(j)}$. For any $\mu \in (0, 1)$, time-sharing between $P_X^{(1)}$ (with a time fraction of $\mu$) and $P_X^{(2)}$ (with a time fraction of $(1 - \mu)$) hence leads to
\begin{eqnarray}
\mu C(D_1) + (1 - \mu) C(D_2) &\leq& \max_{P_X \in \mathcal{P}_{\mu D_1 + (1 - \mu) D_2}} I(X; Y)\nonumber\\
&=& C(\mu D_1 + (1 - \mu) D_2).
\end{eqnarray}
So $C(D)$ is a concave function of $D$.

Utilizing the concavity of $C(D)$, (\ref{eqn:R-bound-c-d}) is further upper bounded by
\begin{eqnarray}
R \leq C\left(\frac{1}{n}\sum_{i = 1}^n \sum_{x\in\mathcal{X}} P_{X_i}(x) d^\ast(x)\right) + P_e^{(n)} R + \frac{1}{n}.
\end{eqnarray}

In order to complete the converse proof, we need to show that for any sequence of $(R, D)$-achievable codes and for all sufficiently large $n$, the following should be satisfied,
\begin{eqnarray}
\label{eqn:estimation-bound}
\frac{1}{n}\sum_{i = 1}^n \sum_{x\in\mathcal{X}} P_{X_i}(x) d^\ast(x) \leq D.
\end{eqnarray}
If (\ref{eqn:estimation-bound}) holds, then we can directly establish the converse since for all sufficiently large $n$,
\begin{eqnarray}
R \leq \frac{1}{n}\sum_{i = 1}^n C(D) + P_e^{(n)} R + \frac{1}{n} \rightarrow C(D),
\end{eqnarray}
with $\lim_{n \rightarrow \infty} P_e^{(n)} = 0$.

{\it Proof of (\ref{eqn:estimation-bound}):} We note that the empirical input distribution $P_{X_i}(x)$ is induced by the uniformly selected message, {\it i.e.}, $X_i = x_i(m)$ with probability $1/|\mathcal{M}|$, for every $m \in \mathcal{M}$. Hence the left hand side of (\ref{eqn:estimation-bound}) can be rewritten as
\begin{eqnarray}
\label{eqn:one-shot-distortion}
\frac{1}{n}\sum_{i = 1}^n \sum_{x\in \mathcal{X}}P_{X_i}(x)d^\ast(x) &=& \frac{1}{n}\sum_{i = 1}^n\frac{1}{|\mathcal{M}|}\sum_{m \in \mathcal{M}} d^\ast(x_i(m))\nonumber\\
&=& \frac{1}{n|\mathcal{M}|} \sum_{m \in \mathcal{M}} \sum_{i = 1}^n \mathbf{E}[d(S_i, h^\ast(X_i, Y_i))|X_i = x_i(m)],
\end{eqnarray}
where the expectation is with respect to $S_i$ and $Y_i$ induced by $x_i(m)$. We rewrite the average distortion definition (\ref{eqn:distortion-def}) as
\begin{eqnarray}
\label{eqn:overall-distortion}
\bar{d}^{(n)} = \frac{1}{n|\mathcal{M}|} \sum_{m \in \mathcal{M}}\sum_{i = 1}^n \mathbf{E}\left[\left. d(S_i, \hat{S}_i) \right| X^n = f_n(m)\right],
\end{eqnarray}
which, for any arbitrarily small $\epsilon > 0$, for all sufficiently large $n$, has to be no greater than $D + \epsilon$ due to the $(R, D)$-achievability requirement. Comparing (\ref{eqn:one-shot-distortion}) and (\ref{eqn:overall-distortion}), it is thus sufficient to show that for each $m \in \mathcal{M}$ and each $1 \leq i \leq n$, in order to prove our desired result,
\begin{eqnarray}
\label{eqn:per-term-dominate}
\mathbf{E}[d(S_i, h^\ast(X_i, Y_i))|X_i = x_i(m)] \leq \mathbf{E}\left[d(S_i, \hat{S}_i)|X^n = f_n(m)\right],
\end{eqnarray}
where the expectation on the right hand side is with respect to $S^n$ and $Y^n$ induced by the transmitted channel inputs $x^n(m)$. To further reduce the problem, we strengthen the channel state estimator by revealing (via a genie) $x_i(m)$ to the estimator, when performing the state estimation. This way, the optimal state estimator for $S_i$, denoted $\tilde{h}^\ast_i(\cdot, \cdot)$, solves
\begin{eqnarray}
\label{eqn:block-estimate}
\min_{\tilde{h}_i: \mathcal{X} \times \mathcal{Y}^n \mapsto \mathcal{S}} \mathbf{E}\left[\left.d\left(S_i, \tilde{h}_i(X_i, Y^n)\right)\right| X_i = x_i(m)\right],
\end{eqnarray}
for every revealed $x_i(m)$, $m \in \mathcal{M}$. In contrast, $h^\ast(\cdot, \cdot)$ solves
\begin{eqnarray}
\min_{h: \mathcal{X}\times \mathcal{Y}\mapsto\mathcal{S}} \mathbf{E}\left[d(S_i, h(X_i, Y_i))|X_i = x_i(m)\right].
\end{eqnarray}
Therefore, (\ref{eqn:per-term-dominate}) can be established by showing that
\begin{eqnarray}
\label{eqn:estimator-compare}
\min_{h: \mathcal{X}\times \mathcal{Y}\mapsto\mathcal{S}} \mathbf{E}\left[d(S_i, h(X_i, Y_i))|X_i = x_i(m)\right] \leq \min_{\tilde{h}_i: \mathcal{X} \times \mathcal{Y}^n \mapsto \mathcal{S}} \mathbf{E}\left[\left.d\left(S_i, \tilde{h}_i(X_i, Y^n)\right)\right|X_i = x_i(m)\right].
\end{eqnarray}
In words, (\ref{eqn:estimator-compare}) indicates that knowing the entire channel output sequence $Y^n$ does not lead to better estimation of $S_i$ versus only knowing $Y_i$. In order to prove (\ref{eqn:estimator-compare}), we need the following lemma.

\begin{lem}
\label{lem:irrelevance}
For three arbitrary random variables $U \in \mathcal{U}$, $V \in \mathcal{V}$, and $W \in \mathcal{W}$, where $W$ is independent of $(U, V)$, and for an arbitrary function $d: \mathcal{U} \times \mathcal{U} \mapsto \mathbb{R}$, we have
\begin{eqnarray}
\min_{f: \mathcal{V} \mapsto \mathcal{U}} \mathbf{E}\left[d\left(U, f(V)\right)\right] = \min_{g: \mathcal{V} \times \mathcal{W} \mapsto \mathcal{U}} \mathbf{E}\left[d\left(U, g(V, W)\right)\right].
\end{eqnarray}
\end{lem}

{\it Proof of Lemma \ref{lem:irrelevance}:} Using the law of total expectation, we have
\begin{eqnarray}
\mathbf{E}\left[d\left(U, g(V, W)\right)\right] = \mathbf{E}\left[\mathbf{E}\left[d\left(U, g(V, W)\right)|W\right]\right],
\end{eqnarray}
where the inner expectation is over $(U, V)$ and the outer expectation is over $W$. Noting that $W$ is independent of $(U, V)$, we have
\begin{eqnarray}
\label{eqn:irrelevance<=}
\min_{g: \mathcal{V} \times \mathcal{W} \mapsto \mathcal{U}} \mathbf{E}\left[d\left(U, g(V, W)\right)\right] &=& \min_{g: \mathcal{V} \times \mathcal{W} \mapsto \mathcal{U}} \mathbf{E}\left[\mathbf{E}\left[d\left(U, g(V, W)\right)|W\right]\right]\nonumber\\
&\geq& \mathbf{E}\left[\min_{g(\cdot, W): \mathcal{V} \mapsto \mathcal{U}} \mathbf{E}\left[d(U, g(V, W))|W\right]\right]\nonumber\\
&=& \mathbf{E}\left[\min_{f: \mathcal{V} \mapsto \mathcal{U}} \mathbf{E}\left[d(U, f(V))\right]\right]\nonumber\\
&=& \min_{f: \mathcal{V} \mapsto \mathcal{U}} \mathbf{E}\left[d(U, f(V))\right].
\end{eqnarray}
On the other hand, since $f: \mathcal{V} \mapsto \mathcal{U}$ is a special form of $g: \mathcal{V} \times \mathcal{W} \mapsto \mathcal{U}$, we have
\begin{eqnarray}
\label{eqn:irrelevance>=}
\min_{f: \mathcal{V} \mapsto \mathcal{U}} \mathbf{E}\left[d\left(U, f(V)\right)\right] \geq \min_{g: \mathcal{V} \times \mathcal{W} \mapsto \mathcal{U}} \mathbf{E}\left[d\left(U, g(V, W)\right)\right].
\end{eqnarray}
Combining (\ref{eqn:irrelevance<=}) and (\ref{eqn:irrelevance>=}) establishes Lemma \ref{lem:irrelevance}.

Now let us apply Lemma \ref{lem:irrelevance} to (\ref{eqn:estimator-compare}). We recognize $S_i$ as $U$, $Y_i$ as $V$, and $(Y_1, \ldots, Y_{i - 1}, Y_{i + 1}, \ldots, Y_n)$ as $W$, and note that with $x_i(m)$ given, $(Y_1, \ldots, Y_{i - 1}, Y_{i + 1}, \ldots, Y_n)$ is independent of $(S_i, Y_i)$ due to the memoryless property of the channel. Therefore Lemma \ref{lem:irrelevance} indicates that (\ref{eqn:estimator-compare}) holds with equality. Consequently, (\ref{eqn:estimation-bound}) holds, thus concluding the converse proof of Theorem \ref{thm:c-d}.

%%%%%%%%%%%%%%%%%%%%%%%%%%%%%%%%%%%%%%%%%%%%%%%%%%%%%%%%%%%%%%%%%%%%%%
\section{Channels with Input Constraints}
\label{sec:continuous}

In this section, we extend Theorem \ref{thm:c-d} to the scenario where besides the state estimation constraint, there also exists an average cost constraint on the channel inputs.

For the basic problem formulation presented in Section \ref{sec:problem-formulation}, we introduce a cost function $v(\cdot): \mathcal{X} \mapsto \mathbb{R}^+\cup \{0\}$, which associates each input letter with a certain nonnegative cost. For a given sequence of inputs $(x_1, \ldots, x_n) \in \mathcal{X}^n$, the resulting total input cost is $\sum_{i = 1}^n v(x_i)$. For an $(|\mathcal{M}|, n)$-code, the average input cost is defined as
\begin{eqnarray}
\bar{v}^{(n)} = \frac{1}{n|\mathcal{M}|} \sum_{m \in \mathcal{M}} \sum_{i = 1}^n v(x_i(m)).
\end{eqnarray}

Subsequently, a tuple $(R, D, V)$ can be used to describe a tradeoff between transmission rate, state estimation distortion, and input cost, which is achievable if there exists a sequence of $(\left\lceil e^{nR} \right\rceil, n)$-codes, indexed by $n = 1, 2, \ldots$, such that $\lim_{n \rightarrow \infty} P_e^{(n)} = 0$, $\limsup_{n \rightarrow \infty} \bar{d}^{(n)} \leq D$, and $\limsup_{n \rightarrow \infty} \bar{v}^{(n)} \leq V$. Therefore, the capacity-distortion-cost function $C(D, V)$ is the supremum of rates $R$ such that $(R, D, V)$ is an achievable tradeoff. Frequently, it is customary to fix $V$, and consider the capacity-distortion function $C(D)$ under that fixed $V$, as we will develop in some examples in Section \ref{sec:example}.

Under such an average input constraint, Theorem \ref{thm:c-d} is extended to the form described by the following theorem.

\begin{thm}
\label{thm:c-d-input}
The capacity-distortion-cost function is
\begin{eqnarray}
C(D, V) = \max_{P_X \in \mathcal{P}_D \cap \mathcal{P}_V} I(X; Y),
\end{eqnarray}
where
\begin{eqnarray}
\mathcal{P}_D &=& \left\{P_X: \sum_{x \in \mathcal{X}} P_X(x) d^\ast(x) \leq D \right\},\\
\mathcal{P}_V &=& \left\{P_X: \sum_{x \in \mathcal{X}} P_X(x) v(x) \leq V \right\}.
\end{eqnarray}
\end{thm}

{\it Proof:} The achievability of $C(D, V)$ follows from that of $C(D)$ as developed in the achievability proof of Theorem \ref{thm:c-d}, combined with consideration of the average input cost constraint, cf. \cite[Sec. 3.4]{kramer08:book}.

To establish the converse part, the argument is as follows. From Theorem \ref{thm:c-d} and the standard capacity-cost result \cite[Chap. 3]{kramer08:book} respectively, we have that any achievable $(R, D, V)$ should satisfy
\begin{eqnarray}
\label{eqn:conv-1}
R \leq \max_{P_X \in \mathcal{P}_D} I(X; Y),
\end{eqnarray}
and
\begin{eqnarray}
\label{eqn:conv-2}
R \leq \max_{P_X \in \mathcal{P}_V} I(X; Y).
\end{eqnarray}
Assume that these exists $R > C(D, V) = \max_{P_X \in \mathcal{P}_D \cap \mathcal{P}_V} I(X; Y)$ such that the tuple $(R, D, V)$ is achievable. Then from (\ref{eqn:conv-1}) and (\ref{eqn:conv-2}), we have either $P_X \notin \mathcal{P}_V$ or $P_X \notin \mathcal{P}_D$, which would in turn violate (\ref{eqn:conv-2}) or (\ref{eqn:conv-1}), respectively. Therefore, no rate $R > C(D, V)$ can be achievable, and the converse of Theorem \ref{thm:c-d-input} is established.

%%%%%%%%%%%%%%%%%%%%%%%%%%%%%%%%%%%%%%%%%%%%%%%%%%%%%%%%%%%%%%%%%%%%%%
\section{Examples}
\label{sec:example}

In this section, we illustrate through examples the capacity-distortion function characterized in the previous sections. The first example examines a simple scenario where the estimation costs are uniform, and specifically shows that for a state-dependent Gaussian channel the capacity-distortion function behaves quite differently than that for the system with the state information at the transmitter. The second example evaluates the capacity-distortion function for certain binary multiplicative channels, and shows that the capacity-distortion function exceeds the tradeoff achieved by training. The third example considers a memoryless Rayleigh fading channel, characterizing its capacity-distortion function within $1.443$ bits ({\it i.e.}, one nat) at high SNR.

\subsection{Channels with Uniform Estimation Costs}

A special case is that $d^\ast(x)$, the estimation cost as defined in (\ref{eqn:d-ast}), is a constant $d_0$ for all $x \in \mathcal{X}$. For this type of channels, the average cost constraint in (\ref{eqn:p-d-set}) exhibits a singular behavior. If $D < d_0$, the joint transmission and state estimation problem is infeasible; otherwise, $\mathcal{P}_D$ consists of all possible input distributions, and thus the capacity-distortion function $C(D)$ is equal to the unconstrained capacity of the channel. One of the simplest channels with uniform estimation costs is the additive channel $Y_i = X_i + S_i + Z_i$, for which as the destination reliably decodes the message, it can subtract $X_i$ from $Y_i$ so that the estimation of $S_i$ becomes independent of $X_i$.

We now briefly contrast our results for the capacity distortion function where the transmitter is oblivious to channel state with the work of \cite{sutivong02:isit, sutivong05:it, kim08:it} wherein the transmitter knows the channel state non-causally. Consider the state-dependent Gaussian channel:
\begin{eqnarray*}
Y_i &=& X_i + S_i + Z_i
\end{eqnarray*}
where $S_i \sim \mathcal{N}(0, Q)$, $Z_i \sim \mathcal{N}(0, N)$ and the transmitted signal has a power constraint of $P$. It is straightforward to show that the capacity distortion function $C(D) = \log \left( 1 + \frac{P}{Q+N} \right)$ for $ D > \frac{QN}{Q +N}$ and zero otherwise for the mean-squared error distortion metric. In contrast, if the transmitter knows the channel state \cite[Thm. 2]{sutivong05:it}, the system can achieve the following tradeoff:
\begin{eqnarray*}
R &\leq& \frac{1}{2} \log\left(1 + \frac{\gamma P}{N}\right),\\
D &\geq& Q\frac{\gamma P + N}{\left(\sqrt{Q} + \sqrt{(1 - \gamma)P}\right)^2 + \gamma P + N},
\end{eqnarray*}
for $0 \leq \gamma \leq 1$. It is clear that channel knowledge at the transmitter enables both an increase in capacity as well as a reduction in distortion of channel state estimation.

\subsection{A Binary Multiplicative Channel}
\label{subsec:bmultiply}

We next consider an example that, while somewhat simple, facilitates drawing insights about the nature of joint transmission and state estimation and employs a distortion metric alternative to the mean-squared error. Consider the following,
\begin{eqnarray}
\underline{Y}_i = S_i \underline{X}_i,
\end{eqnarray}
where $\underline{X}$ and $\underline{Y}$ are length-$K$ blocks so that the super-symbols in the block memoryless channel have alphabets $\mathcal{X}^K = \mathcal{Y}^K = \{0, 1\}^K$ and the multiplication is in the common sense for real numbers. The channel state $S \in \mathcal{S} = \{0, 1\}$ remains fixed for each block, and changes in a memoryless fashion across blocks. We denote $\mathbf{Pr}[S = 1] = r < 1/2$. We adopt the Hamming distance as the distortion measure: $d(s, \hat{s}) = 1$ if and only if $\hat{s} \neq s$ and zero otherwise. We can view $S$ as the status of a jamming source, a fading level, or the status of a primary transmitter in cognitive radio systems. Activating $S$ to its ``effective status'' $S = 0$ essentially shuts down the link between $X$ and $Y$; otherwise, the link from $X$ to $Y$ is noiseless. The tradeoff between communication and channel estimation is straightforward to observe from the nature of the channel: for good estimation of $S$, we want $x= 1$ as often as possible, whereas this would reduce the achieved information rate.

For $K \ge 2$, there are $2^K$ possible vectors for an input super-symbol.  All $\underline{x}$ except for the all-zero $\underline{x} = \underline{0}$ case lead to the same conditional distribution for $\underline{Y}$ as well as the same minimum conditional distortion $d^\ast(\underline{x}) = 0$. From the concavity of mutual information with respect to input distribution, the optimal input distribution should take the following form:
\begin{eqnarray*}
P_{\underline{X}}(\underline{0}) = 1 - p, \quad\mbox{and}\; P_{\underline{X}}(\underline{x}) = p/(2^K - 1), \;\; \forall \underline{x} \neq \underline{0}.
\end{eqnarray*}
We can find that the channel mutual information per channel use is
\begin{eqnarray}
\frac{I(\underline{X}; \underline{Y})}{K} = \frac{1}{K}\left\{H_2(pr) + p\cdot\left[r \log(2^K - 1) - H_2(r)\right] \right\},
\end{eqnarray}
and that the average distortion constraint is
$(1 - p)r \leq D$. The resulting solution for maximizing the mutual information is
\begin{eqnarray*}
&&\mbox{Case 1}:\;\; 2^K > 1 + (1 - r)^{-1/r}:
p^\ast = 1, \; C(D) = \frac{r\log(2^K - 1)}{K} > 0.\\
&&\mbox{Case 2}:\;\; 2^K \leq 1 + (1 - r)^{-1/r}:\\
&&\quad \mbox{if}\;\; D \geq r - \left[1 + \frac{1}{2^K - 1} e^{H_2(r)/r}\right]^{-1} \geq 0:\\
&&\quad p^\ast = \frac{1}{r}\left[1 + \frac{1}{2^K - 1} e^{H_2(r)/r}\right]^{-1}, \; C(D) = \frac{1}{K} \left\{H_2(p^\ast r) + p^\ast \left[r \log(2^K - 1) - H_2(r)\right]\right\};\\
&&\quad \mbox{otherwise}:\\
&&\quad  p^\ast = 1 - \frac{D}{r}, \; C(D) = \frac{1}{K}\left\{H_2(r - D) + \left(1 - \frac{D}{r}\right)\left[r \log(2^K - 1) - H_2(r)\right]\right\}.
\end{eqnarray*}

Case 1 arises because if the channel block length $K$ is sufficiently large such that $2^K > 1 + (1 - r)^{-1/r}$, then the resulting $p^\ast$ as given by Case 2 would be greater than one, which is impossible. In Case 1, we have $P_{\underline{X}}(\underline{0}) = 0$, and all the nonzero symbols are selected with equal probability $1/(2^K - 1)$. In fact, Case 1 kicks in for rather small values of $K$. In our channel model with $r \in [0, 1/2]$, for $r$ smaller than $0.175$, Case 1 arises for $K \geq 2$; and for all $r$ larger than $0.175$, Case 1 arises for $K \geq 3$.

Numerical evaluation of $C(D)$ reveals the trends described above. For relatively large $D$, the average distortion constraint is not active, and thus the optimal input distribution coincides with that for the unconstrained channel capacity. As the estimation distortion constraint $D$ falls below a threshold, the average distortion constraint becomes active, and the capacity-distortion function $C(D)$ decreases from the unconstrained channel capacity.

For $K=1$, we can show that as $D \rightarrow 0$,
\begin{eqnarray}
\label{eqn:smultiply-smallD}
C(D) = \frac{\log(1 - r)}{-r} D + o(D)
\end{eqnarray}
which reflects a linear increase in capacity as we loosen the distortion requirement. For $K > 1$, we have
\begin{eqnarray}
C(0) = \frac{r\log(2^K - 1)}{K} > 0.
\end{eqnarray}
For comparison, let us consider a suboptimal approach based upon training; the source transmits $X = 1$ in the first channel use in each channel block. The receiver can thus perfectly estimate the channel state $S$ and achieve $D = 0$. The encoder then can use the remaining $(K - 1)$ channel uses in each channel block to encode information, and the resulting achievable rate is
\begin{eqnarray}
R(0) = \frac{r \log(2^{K - 1})}{K}.
\end{eqnarray}
Comparing $C(0)$ and $R(0)$, we notice that their ratio approaches one as $K \rightarrow \infty$, consistent with the intuition that training usually leads to negligible rate loss for channels with long coherence blocks.  %as proved in \cite{McElieceStark}.
But for small coherence blocks, the joint approach outperforms the training based approach.

\subsection{Memoryless Rayleigh Fading Channel}

Consider a discrete-time memoryless Rayleigh fading channel with scalar input and output, as
\begin{eqnarray}
\label{eqn:rayleigh-model}
Y = S X + Z,
\end{eqnarray}
where $X \in \mathbb{C}$ is the channel input, and $Y \in \mathbb{C}$ is the channel output. There is an average power constraint on $X$, as
\begin{eqnarray}
\mathbf{E}[|X|^2] \leq \rho.
\end{eqnarray}
The fading coefficient $S \in \mathbb{C}$ is the channel state to estimate, following a zero-mean unit-variance circular complex Gaussian distribution, $\mathcal{CN}(0, 1)$. The additive noise $Z \in \mathbb{C}$ is also $\mathcal{CN}(0, 1)$. The distortion function is quadratic, {\it i.e.},
\begin{eqnarray}
d(s, \hat{s}) = |s - \hat{s}|^2.
\end{eqnarray}
Therefore, the optimal one-shot estimator $h^\ast(x, y)$ is the minimum-mean-squared-error (MMSE) estimator, as
\begin{eqnarray}
h^\ast(x, Y) = \frac{x^\dag}{|x|^2 + 1} Y,
\end{eqnarray}
and the resulting estimation cost $d^\ast(x)$ is the MMSE
\begin{eqnarray}
d^\ast(x) = \frac{1}{|x|^2 + 1}.
\end{eqnarray}
So the capacity-distortion function $C(D)$ is characterized by the following optimization:
\begin{eqnarray}
\label{eqn:rayleigh-problem}
&&\max_{dP_X} I(X; Y),\\
\mbox{s.t.}&&\quad \int_{\mathbb{C}} |x|^2 dP_X(x) \leq \rho,\nonumber\\
&&\int_{\mathbb{C}} \frac{1}{|x|^2 + 1} dP_X(x) \leq D.\nonumber
\end{eqnarray}

Even without the channel state estimation constraint (that is, $D \geq 1$), neither the capacity nor the capacity-achieving input distribution of (\ref{eqn:rayleigh-problem}) is fully known. It has been proved in \cite{abou01:it} that the (power-constrained) channel capacity is achieved by a discrete input distribution with a finite number of mass points, including a mass point at $X = 0$. For the high-SNR regime, it is also known that the channel capacity grows double-logarithmically, {\it i.e.}, $C = \mathcal{O}(\log\log \rho)$ as $\rho \rightarrow \infty$ \cite{taricco97:el}. More precisely, it is established in \cite{lapidoth03:it} that for fairly general non-coherent fading channels, $C = \log\log \rho + \chi + o(1)$ as $\rho \rightarrow \infty$, where $\chi$ is a constant and is called the fading number. For the scalar memoryless Rayleigh fading channel considered here, the fading number $\chi = - 1 - \gamma$ where $\gamma = 0.5772...$ is Euler's constant.

For the $C(D)$ optimization problem (\ref{eqn:rayleigh-problem}), we note that the two constraints have conflicting effects on the distribution of $X$. The average power constraint tends to ``stretch'' the support set of $X$ toward zero because otherwise a certain amount of input power would be wasted; in contrast, the channel state estimation constraint tends to ``push'' the support set of $X$ away from zero because otherwise the average distortion may violate the constraint. We focus on the high-SNR regime with $\rho$ growing without bound, in which it is possible to simultaneously achieve large (increasing without bound as $\rho \rightarrow \infty$) transmission rate and small (decreasing toward zero as $\rho \rightarrow \infty$) estimation distortion. The following theorem characterizes some asymptotic behaviors of $C(D)$ as $\rho \rightarrow \infty$.
\begin{thm}
\label{thm:rayleigh-high-snr}
For the discrete-time memoryless Rayleigh fading channel (\ref{eqn:rayleigh-model}) with average power constraint $\rho$ and channel state estimation constraint $D$:
\begin{enumerate}
\item If $\lim_{\rho \rightarrow \infty} D\rho^\alpha = \kappa$, where $0 \leq \alpha < 1$ and $0 < \kappa < \infty$ are both constants, for sufficiently large $\rho$, $C(D)$ satisfies
\begin{eqnarray}
\log\log \rho + \log(1 - \alpha) - 1 - \gamma \leq C(D) \leq \log\log \rho + \log(1 - \alpha) - \gamma.
\end{eqnarray}
\item If $\lim_{\rho \rightarrow \infty} D\rho = \kappa < \infty$, then $C(D)$ does not grow to infinity for all $\rho$.
\end{enumerate}
\end{thm}

The proof of Theorem \ref{thm:rayleigh-high-snr} is in Appendix, and is based on an induced additive-noise model for the memoryless Rayleigh fading channel introduced in \cite{zhang06:it}.

%%%%%%%%%%%%%%%%%%%%%%%%%%%%%%%%%%%%%%%%%%%%%%%%%%%%%%%%%%%%%%%%%%%%%%
\section{Extension to Two-User MAC}
\label{sec:mac}

In this section, we extend the transmission versus state estimation problem to two-user MAC with state estimation, establishing its capacity-distortion region. We consider a two-user discrete memoryless MAC with state, whose channel transition probability distribution is described by $P(y|x_1, x_2, s)$, where $x_1 \in \mathcal{X}_1$ and $x_2 \in \mathcal{X}_2$ are the input alphabets for the first and second sources, respectively. The channel state $S$ is a random variable with PMF $P_S(s)$ over the state alphabet $\mathcal{S}$. The channel output alphabet is $\mathcal{Y}$.

We consider an $(|\mathcal{M}_1|, |\mathcal{M}_2|, n)$-code, which consists of two encoders $f_{1, n}: \mathcal{M}_1 \mapsto \mathcal{X}_1^n$ and $f_{2, n}: \mathcal{M}_2 \mapsto \mathcal{X}_2^n$, a decoder $g_n: \mathcal{Y}^n \mapsto \mathcal{M}_1 \times \mathcal{M}_2$, and a state estimator $h_n: \mathcal{Y}^n \mapsto \mathcal{S}^n$. Due to the presence of two sources, we define the average probability of decoding error as
\begin{eqnarray}
P_e^{(n)} = \frac{1}{|\mathcal{M}_1| |\mathcal{M}_2|} \sum_{(m_1, m_2) \in \mathcal{M}_1 \times \mathcal{M}_2} \mathbf{Pr}[g_n(Y^n) \neq (m_1, m_2)|X_1^n = f_{1, n}(m_1), X_2^n = f_{2, n}(m_2)],
\end{eqnarray}
and the average distortion of state estimation as
\begin{eqnarray}
\bar{d}^{(n)} = \frac{1}{|\mathcal{M}_1||\mathcal{M}_2|} \sum_{(m_1, m_2) \in \mathcal{M}_1 \times \mathcal{M}_2} \mathbf{E}\left[\left.\frac{1}{n}\sum_{i = 1}^n d(S_i, \hat{S}_i)\right| X_1^n = f_{1, n}(m_1), X_2^n = f_{2, n}(m_2)\right].
\end{eqnarray}
We say that a tuple $(R_1, R_2, D)$ is achievable if there exists a sequence of $(\left\lceil e^{nR_1} \right\rceil, \left\lceil e^{nR_2} \right\rceil, n)$-codes, indexed by $n = 1, 2, \ldots$, such that $\lim_{n \rightarrow \infty} P_e^{(n)} = 0$, and $\limsup_{n \rightarrow \infty} \bar{d}^{(n)} \leq D$, and define the capacity-distortion region $\mathcal{C}(D)$ as the closure of rate-pairs $(R_1, R_2)$ such that $(R_1, R_2, D)$ is an achievable transmission-state estimation tradeoff.

Analogous to the single-user case, we define the minimal conditional distortion, or, estimation cost, for the two-user MAC as
\begin{eqnarray}
d^\ast(x_1, x_2) = \min_{h: \mathcal{X}_1 \times \mathcal{X}_2 \times \mathcal{Y} \mapsto \mathcal{S}} \mathbf{E}[d(S, h(X_1, X_2, Y))| X_1 = x_1, X_2 = x_2],
\end{eqnarray}
for $(x_1, x_2) \in \mathcal{X}_1 \times \mathcal{X}_2$.

Combining the proofs of Theorem \ref{thm:c-d} and the standard MAC coding theorem (see, {\it e.g.}, \cite[Thm. 15.3.1]{cover06:book}), we have the following theorem characterizing the capacity-distortion region.
\begin{thm}
\label{thm:c-d-mac}
For the two-user state-dependent MAC, its capacity-distortion region $\mathcal{C}(D)$ is the union of all $(R_1, R_2)$ satisfying
\begin{eqnarray}
R_1 &\leq& I(X_1; Y|X_2, Q),\\
R_2 &\leq& I(X_2; Y|X_1, Q),\\
R_1 + R_2 &\leq& I(X_1, X_2; Y|Q),
\end{eqnarray}
over product distributions $P_Q(q)P_{X_1|Q}(x_1|q)P_{X_2|Q}(x_2|q)P_{Y|X_1, X_2}(y|x_1, x_2)$ on $\mathcal{Q}\times \mathcal{X}_1\times \mathcal{X}_2 \times \mathcal{Y}$ satisfying
\begin{eqnarray}
\sum_{(q, x_1, x_2) \in \mathcal{Q}\times \mathcal{X}_1\times \mathcal{X}_2} P_Q(q)P_{X_1|Q}(x_1|q)P_{X_2|Q}(x_2|q)d^\ast(x_1, x_2) \leq D.
\end{eqnarray}
Here the cardinality $|\mathcal{Q}| \leq 5$.
\end{thm}

{\it Proof:} The achievability part follows from the standard MAC capacity theorem \cite[Sec. 15.3.1 and Thm. 15.3.4]{cover06:book} based on random codebooks and typicality decoding, combined with the distortion bounding procedure in Section \ref{sec:achievability} using the asymptotically reliable inputs sequence $(\hat{X}_1^n, \hat{X}_2^n)$ to estimate the channel states.

To establish the converse, we begin by following the same bounding steps as in the converse proof of standard MAC capacity theorem (cf. \cite[Sec. 15.3.4]{cover06:book}). Considering any sequence of $(\left\lceil e^{nR_1} \right\rceil, \left\lceil e^{nR_2} \right\rceil, n)$-codes with $\lim_{n \rightarrow \infty} P_e^{(n)} = 0$, the bounding procedure arrives at
\begin{eqnarray}
\label{eqn:mac-r1-noQ}
R_1 &\leq& \frac{1}{n} \sum_{i = 1}^n I(X_{1, i}; Y_i|X_{2, i}) + \epsilon_n,\\
\label{eqn:mac-r2-noQ}
R_2 &\leq& \frac{1}{n} \sum_{i = 1}^n I(X_{2, i}; Y_i|X_{1, i}) + \epsilon_n,\\
\label{eqn:mac-r12-noQ}
R_1 + R_2 &\leq& \frac{1}{n} \sum_{i = 1}^n I(X_{1, i}, X_{2, i}; Y_i) + \epsilon_n,
\end{eqnarray}
where $\lim_{n \rightarrow \infty} \epsilon_n = 0$.

Since the considered sequence of $(\left\lceil e^{nR_1} \right\rceil, \left\lceil e^{nR_2} \right\rceil, n)$-codes also needs to satisfy the state estimation distortion constraint, we have that the induced average distortion must not exceed $D + \epsilon_n$, {\it i.e.},
\begin{eqnarray}
\label{eqn:distortion-bound-mac-1}
\bar{d}^{(n)} = \frac{1}{n |\mathcal{M}_1||\mathcal{M}_2|} \sum_{(m_1, m_2)\in \mathcal{M}_1 \times \mathcal{M}_2} \sum_{i = 1}^n \mathbf{E}\left[d(S_i, \hat{S}_i) | X_1^n = f_{1, n}(m_1), X_2^n = f_{2, n}(m_2)\right] \leq D + \epsilon_n.
\end{eqnarray}
Using Lemma \ref{lem:irrelevance}, as in the converse proof for the single-user case in Section \ref{sec:converse}, we have from (\ref{eqn:distortion-bound-mac-1}) that for any given sequence of $(\left\lceil e^{nR_1} \right\rceil, \left\lceil e^{nR_2} \right\rceil, n)$-codes, it is necessary to have
\begin{eqnarray}
\frac{1}{n|\mathcal{M}_1||\mathcal{M}_2|} \sum_{(m_1, m_2)\in \mathcal{M}_1 \times \mathcal{M}_2} \sum_{i = 1}^n \mathbf{E}\left[d(S_i, h^\ast(X_{1, i}, X_{2, i}, Y_i)) | X_{1, i} = x_{1, i}(m_1), X_{2, i} = x_{2, i}(m_2)\right] \leq D + \epsilon_n,
\end{eqnarray}
for all sufficiently large $n$. That is (cf. (\ref{eqn:one-shot-distortion})),
\begin{eqnarray}
\label{eqn:mac-d-noQ}
\frac{1}{n} \sum_{i = 1}^n \sum_{x_1 \in \mathcal{X}_1} \sum_{x_2 \in \mathcal{X}_2} P_{X_{1, i}}(x_1) P_{X_{2, i}}(x_2) d^\ast(x_1, x_2) \leq D + \epsilon_n,
\end{eqnarray}
where we use the fact that the two encoders are independent.

Now, as we let $n$ grow without bound and introduce a uniform random variable $Q$ over $\{1, 2, \ldots, n\}$, following the same argument as in \cite[Sec. 15.3.4]{cover06:book}, the region of $(R_1, R_2, D)$ described by (\ref{eqn:mac-r1-noQ})-(\ref{eqn:mac-r12-noQ}) and (\ref{eqn:mac-d-noQ}) can be equivalently rewritten as
\begin{eqnarray}
\label{eqn:mac-r1}
R_1 &\leq& I(X_1; Y|X_2, Q),\\
\label{eqn:mac-r2}
R_2 &\leq& I(X_2; Y|X_1, Q),\\
\label{eqn:mac-r12}
R_1 + R_2 &\leq& I(X_1, X_2; Y|Q),\\
\label{eqn:mac-d}
D &\geq& \sum_{(q, x_1, x_2) \in \mathcal{Q} \times \mathcal{X}_1 \times \mathcal{X}_2} P_Q(q) P_{X_1|Q}(x_1|q) P_{X_2|Q}(x_2|q) d^\ast(x_1, x_2).
\end{eqnarray}

To conclude the proof, we use Carath\'{e}odory's theorem to bound the cardinality of $Q$, as that for the standard MAC capacity theorem in \cite[Sec. 15.3.3]{cover06:book}. The region described by (\ref{eqn:mac-r1})-(\ref{eqn:mac-d}) define a connected compact set in four dimensions, and hence we can restrict the cardinality of $Q$ to at most $5$ in the capacity-distortion region. Theorem \ref{thm:c-d-mac} thus is established.

%%%%%%%%%%%%%%%%%%%%%%%%%%%%%%%%%%%%%%%%%%%%%%%%%%%%%%%%%%%%%%%%%%%%%%
\section{Conclusions}
\label{sec:conclusion}

In this paper, we introduced a joint information transmission and channel state estimation problem for state-dependent channels, and characterized its fundamental tradeoff by formulating it as a channel coding problem with input distribution constrained by an average estimation cost constraint. Key to our problem formulation is the assumption that the transmitter is oblivious to the channel state information. The resulting capacity-distortion function permits a systematic investigation of the channel's capability for transmission and state estimation. We showed that non-coherent communication coupled with channel state estimation conditioned on treating the decoded message as training achieves the capacity-distortion function. We extended our results to multiple access channels, which leads to a coupled cost constraint on the input distributions for the transmitting sources. Future research topics include specializing the general framework to particular channel models in realistic applications, and generalizing the results to multiuser systems and channels with generally correlated state processes.

%%%%%%%%%%%%%%%%%%%%%%%%%%%%%%%%%%%%%%%%%%%%%%%%%%%%%%%%%%%%%%%%%%%%%%
\section*{Acknowledgement}

The authors would like to thank the anonymous reviewers for their comments which have significantly improved the paper, and Gerhard Kramer for extensive discussions about the MAC capacity region versus distortion problem.

%%%%%%%%%%%%%%%%%%%%%%%%%%%%%%%%%%%%%%%%%%%%%%%%%%%%%%%%%%%%%%%%%%%%%%
\section*{Appendix}
\label{sec:appendix}

\subsection{Proof of Theorem \ref{thm:rayleigh-high-snr}}
\label{subsec:proof-rayleigh}

Following the development in \cite[Sec. II]{zhang06:it}, the mutual information $I(X; Y)$ is equal to another mutual information $I(U; T)$, where $U$ and $T$ are the channel input and output of an additive-noise channel
\begin{eqnarray}
\label{eqn:additive-model}
T = U + W,
\end{eqnarray}
with $U = (1/2)\log \left(|X|^2 + 1\right)$ and $T = \log |Y|$. The additive noise $W$ is independent of $U$, and has PDF
\begin{eqnarray}
f_W(w) = 2\exp[2w - \exp(2w)], \quad w \in (-\infty, \infty).
\end{eqnarray}
Accordingly, the two constraints in (\ref{eqn:rayleigh-problem}) can be equivalently rewritten in terms of $U$, and the optimization problem becomes
\begin{eqnarray}
\label{eqn:rayleigh-problem-induced}
&&\max_{dP_U} I(U; T),\\
\mbox{s.t.}&&\quad \int_0^\infty e^{2u} dP_U(u) \leq \rho + 1\nonumber\\
&&\int_0^\infty e^{-2u} dP_U(u) \leq D.\nonumber
\end{eqnarray}

\subsubsection{Lower Bound of $C(D)$}

A lower bound of $I(U; T)$ is given by \cite[Eqn. (17)]{zhang06:it}
\begin{eqnarray}
I(U; T) \geq h(U) + \log\sqrt{1 + e^{-2[h(U) - h(W)]}} - h(W).
\end{eqnarray}
Consider a continuous distribution of $U$ with the following PDF,
\begin{eqnarray}
p_U(u) = 1/\Delta \quad \mbox{for}\; u \in [\underline{u}, \underline{u} + \Delta],
\end{eqnarray}
and zero otherwise. Furthermore, let both constraints in (\ref{eqn:rayleigh-problem-induced}) be active, namely,
\begin{eqnarray}
\label{eqn:constraint-1}
\frac{1}{\Delta} \int_{\underline{u}}^{\underline{u} + \Delta} e^{2u}du = \rho + 1,\\
\label{eqn:constraint-2}
\frac{1}{\Delta} \int_{\underline{u}}^{\underline{u} + \Delta} e^{-2u}du = D.
\end{eqnarray}
Such a uniform distribution of $U$ thus leads to a lower bound on $C(D)$ as
\begin{eqnarray}
C(D) \geq \log \Delta + \log \sqrt{1 + e^{-2[\log \Delta - 1 + \log 2 - \gamma]}} - 1 + \log 2 - \gamma,
\end{eqnarray}
where we note that $h(W) = 1 - \log 2 + \gamma$ \cite[Lem. 2.1]{zhang06:it}. Hence in order to characterize the asymptotic behavior of the lower bound, we only need to investigate how $\Delta$ scales with $\rho$ and $D$. To this end, we solve (\ref{eqn:constraint-1}) and (\ref{eqn:constraint-2}) in parallel, to get
\begin{eqnarray}
\label{eqn:Delta-relationship}
e^{2\Delta} = 2 \Delta^2 (\rho + 1) D \left[1 + \sqrt{1 + \frac{1}{\Delta^2 (\rho + 1) D}}\right] + 1.
\end{eqnarray}

The right hand side of (\ref{eqn:Delta-relationship}) is a monotone increasing function of $\Delta^2 (\rho + 1) D$ over $(0, \infty)$, increasing from one to infinity. Consider the scaling of $\rho \rightarrow \infty$ and $\lim_{\rho \rightarrow \infty} D\rho^\alpha = \kappa$, where $0 \leq \alpha < 1$ and $0 < \kappa < \infty$ are both constants. There are only two possibilities in such an asymptotic regime: $\Delta \rightarrow 0$ or $\Delta \rightarrow \infty$. If $\Delta \rightarrow 0$, then from (\ref{eqn:constraint-1}) it follows that $e^{2\underline{u}} \approx (\rho + 1)$, which, when combined with (\ref{eqn:constraint-2}), leads to $(\rho + 1)D \approx 1$. But this is in contradiction with the assumption that $\rho D \approx \kappa \rho^{1 - \alpha} \rightarrow \infty$. So the only possibility is $\Delta \rightarrow \infty$, and from (\ref{eqn:Delta-relationship}) we can further bound $\Delta$ through
\begin{eqnarray}
e^{2\Delta} > 4 \Delta^2 (\rho + 1) D > 4 \rho D,
\end{eqnarray}
which leads to
\begin{eqnarray}
\frac{e^{2\Delta}}{\rho^{1 - \alpha}} > 4 D \rho^\alpha \rightarrow 4 \kappa.
\end{eqnarray}
Consequently, we have for sufficiently large $\rho$,
\begin{eqnarray}
\log \Delta > \log\log \rho + \log(1 - \alpha) - \log 2,
\end{eqnarray}
which leads to
\begin{eqnarray}
C(D) \geq \log\log \rho + \log(1 - \alpha) - 1 - \gamma.
\end{eqnarray}

On the other hand, if $\lim_{\rho \rightarrow \infty} D\rho = \kappa$ ({\it i.e.}, $\alpha = 1$), then from (\ref{eqn:Delta-relationship}) it is apparent that $\Delta$ does not grow without bound as $\rho \rightarrow \infty$, and consequently the lower bound of $C(D)$ is finite.

\subsubsection{Upper Bound of $C(D)$}

From the additive-noise channel model (\ref{eqn:additive-model}), We have
\begin{eqnarray}
I(U; T) = h(T) - h(T|U) = h(T) - h(W) = h(T) - 1 + \log 2 - \gamma.
\end{eqnarray}
Therefore, an upper bound of $C(D)$ is obtained by upper bounding $h(T)$.

First, since $e^{-2u}$ is a convex function, from Jensen's inequality, the second constraint in (\ref{eqn:rayleigh-problem-induced}) leads to
\begin{eqnarray}
&&D \geq \int_0^\infty e^{-2u} dP_U(u) \geq \exp\left[-2\int_0^\infty u dP_U(u)\right],\nonumber\\
&&i.e., \quad \int_0^\infty u dP_U(u) \geq \frac{1}{2}\log \frac{1}{D}.
\end{eqnarray}
This leads to a constraint on the expectation of the additive-noise channel output, as
\begin{eqnarray}
\mathbf{E}[T] = \mathbf{E}[U] + \mathbf{E}[W] \geq \frac{1}{2}\log \frac{1}{D},
\end{eqnarray}
where we note that $\mathbf{E}[W] = 0$ \cite[Lem. 2.1]{zhang06:it}. Meanwhile, the additive-noise channel output satisfies another constraint as
\begin{eqnarray}
\mathbf{E}[\exp(2T)] = \mathbf{E}[|Y|^2] = \rho + 1.
\end{eqnarray}
So we can upper bound $h(T)$ by solving the following maximum-entropy problem:
\begin{eqnarray}
\label{eqn:max-entropy-T}
&&\max h(T)\\
\mbox{s.t.}&&\quad \mathbf{E}[T] = \Lambda \geq \frac{1}{2}\log \frac{1}{D},\nonumber\\
&& \mathbf{E}[\exp(2T)] = \rho + 1.\nonumber
\end{eqnarray}
The solution to (\ref{eqn:max-entropy-T}) is (cf. \cite[App. A]{zhang06:it})
\begin{eqnarray}
p_T^\ast(t) = 2 \left(\frac{\mu}{\rho + 1}\right)^\mu \frac{1}{\Gamma(\mu)} \exp\left[2\mu t - \frac{\mu}{\rho + 1}\exp(2t)\right],
\end{eqnarray}
where $\Gamma(\cdot)$ is the Gamma function defined as $\Gamma(z) = \int_0^\infty t^{z - 1} e^{-t} dt$ for $z \in \mathbb{C}$ with $\mathbb{Re} z > 0$, $\mu > 0$ is determined by the equation
\begin{eqnarray}
\label{eqn:mu-relationship}
\log \mu - \psi(\mu) = \log(\rho + 1) - 2\Lambda
\end{eqnarray}
and $\psi(\mu) = \frac{d}{d\mu}\log\Gamma(\mu)$ which is the Psi function (also known as the digamma function) \cite{gradshteyn94:book}. Such a maximum-entropy PDF of $p_T^\ast(t)$ leads to an upper bound of $C(D)$ as
\begin{eqnarray}
\label{eqn:cd-ub}
C(D) \leq \log \Gamma(\mu) - \mu \psi(\mu) + \mu - \gamma - 1.
\end{eqnarray}
To tighten the upper bound, we first notice that the right-hand side of (\ref{eqn:cd-ub}) is increasing with $\mu > 0$, so that the tightest upper bound is obtained when $\mu$ is minimized. We then notice that the left-hand side of (\ref{eqn:mu-relationship}) is decreasing with $\mu > 0$, so that the minimum allowed $\mu$ is attained when $\Lambda = (1/2)\log \frac{1}{D}$, and we rewrite (\ref{eqn:mu-relationship}) as
\begin{eqnarray}
\label{eqn:mu-relationship-1}
\log \mu - \psi(\mu) = \log (\rho + 1)D.
\end{eqnarray}

Now, consider the scaling of $\rho \rightarrow \infty$ and $\lim_{\rho \rightarrow \infty} D\rho^\alpha = \kappa$, where $0 \leq \alpha < 1$ and $0 < \kappa < \infty$ are both constants. The right-hand side of (\ref{eqn:mu-relationship-1}) hence scales like $(1 - \alpha)\log \rho + \log \kappa + o(1)$, and we need to enforce $\mu \rightarrow 0$ as $\rho \rightarrow \infty$. More precisely, by noting that $\psi(\mu) = -1/\mu - \gamma + \pi^2\mu/6 - \ldots$ for $\mu \approx 0$, we can write the left-hand side of (\ref{eqn:mu-relationship-1}) as $1/\mu - \log(1/\mu) + \gamma + o(1)$. Comparing the two sides yields $1/\mu = (1 - \alpha)\log\rho + O(1)$. On the other hand, as $\mu \rightarrow 0$, the upper bound of $C(D)$ (\ref{eqn:cd-ub}) scales like
\begin{eqnarray}
\label{eqn:cd-ub-1}
C(D) \leq \log (1/\mu) - \gamma + o(1).
\end{eqnarray}
Substituting $1/\mu = (1 - \alpha)\log\rho + O(1)$ into (\ref{eqn:cd-ub-1}), we finally reach
\begin{eqnarray}
C(D) \leq \log\log\rho + \log(1 - \alpha) - \gamma + o(1).
\end{eqnarray}

Finally, if $\lim_{\rho \rightarrow \infty} D\rho = \kappa$ ({\it i.e.}, $\alpha = 1$), then (\ref{eqn:mu-relationship-1}) approaches
\begin{eqnarray}
\log \mu - \psi(\mu) = \log \kappa
\end{eqnarray}
as $\rho \rightarrow \infty$, whose solution $\mu$ is finite and bounded away from zero. Consequently, the upper bound of $C(D)$ is finite.

%%%%%%%%%%%%%%%%%%%%%%%%%%%%%%%%%%%%%%%%%%%%%%%%%%%%%%%%%%%%%%%%%%%%%%
%\bibliographystyle{ieee}
%\bibliography{jcs-rev3_um}

%%%%%%%%%%%%%%%%%%%%%%%%%%%%%%%%%%%%%%%%%%%%%%%%%%%%%%%%%%%%%%%%%%%%%%%%%%%%%%%%%%%%%%%%%%%%%%%%%%%%%%%%%%%%
{\bf Authors' Short Bio:}

{\bf Wenyi Zhang} (S'00, M'07, SM'11) was born in Chengdu, Sichuan, China in 1979. He received the B.E. degree in automation from Tsinghua University, Beijing, China, in 2001, and the M.S. and Ph.D. degrees in electrical engineering both from the University of Notre Dame, Notre Dame, IN, in 2003 and 2006, respectively. From September 2006, he was affiliated with the Communication Sciences Institute (CSI), University of Southern California (USC), Los Angeles, as a Postdoctoral Research Associate. From May 2008, he was a Senior System Engineer of the Qualcomm Corporate Research and Development, Qualcomm Inc., San Diego, CA. Since 2010, he has been with the faculty of Department of Electronic Engineering and Information Science, University of Science and Technology of China. Dr. Zhang was awarded the Program for New Century Excellent Talents in University (NCET) from the Ministry of Education of P.R.C in 2009, and was selected as candidate for 100 Talents Program of Chinese Academy of Sciences in 2010. Dr. Zhang is a senior member of IEEE, member of SIAM and Sigma Xi. His research interests include communication theory and wireless communications, theory and practice of cognitive radio, information theory, and statistical signal processing and applications.

{\bf Satish Vedantam} received the Bachelor of Technology degree from Indian Institute of Technology, Madras in 2001. He also received Masters degrees in Electrical Engineering (2003) and Applied Mathematics (2007) and a Ph.D. in Electrical Engineering (2009) all from the University of Southern California. While at USC, he was a recipient of the Annenberg fellowship from 2007 to 2008. He worked as a research analyst with Citigroup Global Markets Inc. from 2003 to 2005 and has been working in the research and development department at Bloomberg LP since 2009. His recent interests include crossover areas between mathematics and engineering with a special emphasis on problems involving large data sets.

{\bf Urbashi Mitra} (S'88, M'88, SM'04, F'08) received the B.S. and the M.S. degrees from the University of California at Berkeley in 1987 and 1989 respectively, both in Electrical Engineering and Computer Science. From 1989 until 1990 she worked as a Member of Technical Staff at Bellcore in Red Bank, NJ. In 1994, she received her Ph.D. from Princeton University in Electrical Engineering. From 1994 to 2000, Dr. Mitra was a member of the faculty of the Department of Electrical Engineering at The Ohio State University, Columbus, Ohio. In 2001, she joined the Department of Electrical Engineering at the University of Southern California, Los Angeles, where she is currently a Professor. Dr. Mitra is currently an Associate Editor for the IEEE the Journal of Oceanic Engineering. She was previously an Associate Editor for the IEEE Transactions on Communications and the Transactions on Information Theory. Dr. Mitra served two terms as a member of the IEEE Information Theory Society's Board of Governors (2002-2007). She is the recipient of: Best Applications Paper Award - 2009 International Conference on Distributed Computing in Sensor Systems, the Viterbi School of Engineering Dean's Faculty Service Award (2009), USC Mellon Mentoring Award (2008), IEEE Fellow (2007), Texas Instruments Visiting  Professor (Fall 2002, Rice University), 2001 Okawa Foundation Award, 2000 Lumley Award for Research (OSU College of Engineering), 1997 MacQuigg Award for Teaching (OSU College of Engineering), 1996 National Science Foundation (NSF) CAREER Award, and a 1994 NSF International Post-doctoral Fellowship. She has co-chaired the IEEE Communication Theory Symposium at ICC 2003 in Anchorage, AK and the first ACM Workshop on Underwater Networks at Mobicom 2006, Los Angeles, CA. Dr. Mitra was the tutorials Chair for IEEE ISIT 2007 in Nice, France and the Finance Chair for IEEE ICASSP 2008 in Las Vegas, NV. Dr. Mitra has held visiting appointments at: the Technical University of Delft, Stanford University, Rice Unviersity, and the Eurecom Institute. She served as co-Director of the Communication Sciences Institute at the University of Southern California from 2004-2007.

\end{document}